# Space-Time Gradient Metasurfaces

Y. Hadad D. L. Sounas and A. Alu

Department of Electrical and Computer Engineering, The University of Texas at Austin
1616 Guadalupe St., UTA 7.215, Austin, TX 78701, USA
alu@mail.utexas.edu

**Metasurfaces characterized by a transverse gradient of local impedance have recently opened exciting directions for light manipulation at the nanoscale. Here we add a temporal gradient to the picture, showing that spatio-temporal variations over a surface may largely extend the degree of light manipulation in metasurfaces, and break several of their constraints associated to symmetries. As an example, we synthesize a non-reciprocal classical analogue to electromagnetic induced transparency, opening a narrow window of one-way transmission in an otherwise opaque surface. These properties pave the way to magnetic-free, planarized non-reciprocal ultrathin surfaces for free-space isolation.**

Snell's law of reflection and refraction describes the fact that at the interface between two homogeneous media the wave momentum is conserved. Transversely inhomogeneous frequency-selective surfaces at radio-frequencies and gradient optical metasurfaces have been recently proposed to bypass the conventional form of Snell's law by introducing clever transverse spatial modulations that can add an abrupt additional momentum discontinuity to the incident wave, yielding unusual scattering responses and 'generalized refraction laws' over a surface [1]-[16]. While these concepts have opened a plethora of interesting possibilities for physicists and engineers, allowing manipulation of light over a thin surface, there are still fundamental constraints that a gradient metasurface cannot overcome. For instance, a thin electric surface is inherently limited in the amount of energy that it can couple into an anomalously refracted beam due to geometrical symmetries [2], requiring the use of thicker geometries or stacks.

Another fundamental constraint that gradient metasurfaces have to comply with is associated with reciprocity and time-reversal symmetry, which requires

$$T(\theta_1,\theta_2) = T(\theta_2,\theta_1), \ R(\theta_1,\theta_2) = R(\theta_2,\theta_1), \qquad (1)$$

where $R(\theta_1,\theta_2)$ and $T(\theta_1,\theta_2)$ are the reflection and transmission coefficients for a plane wave that hits at angle $\theta_1$ to a plane wave that is reflected or transmitted at angle $\theta_2$. Eq. (1) implies that, if we are able to transmit energy through the surface at a particular angle and refract it towards a specific direction, the plane wave with same transverse momentum coming back from the other side should be able to couple as well to the original plane wave. In the same way, the plane wave hitting at the complementary angle should be transmitted with equal intensity, as illustrated in Fig 1a. These constraints may be overcome only by breaking time-reversal symmetry, which is possible using magneto-optical effects[17], nonlinearities [18], or spatiotemporal modulation [19]-[21] and moving media [22]-[25]. Magneto-optical effects require bulky magnets and are difficultly accessible at optical frequencies, while nonlinearities are power dependent and require electrically large volumes. In addition, all the previously reported solutions do not allow full transmission, and are limited to usual refraction/reflection laws, being formed by transversely homogeneous or quasi-homogeneous surfaces.

In this letter we show the possibility to overcome these symmetry-related limitations of conventional spatially gradient metasurfaces by adding transverse temporal gradients. For the sake of clarity and mathematical tractability, we restrict ourselves to the simplest gradient impedance surface – a sinusoidally modulated impedance. However, we emphasize that the results that we develop here are extendable to any type of surface gradients, as sophisticated as in [1]-[14].

By combining the concept of temporal and spatial gradients in ultrathin metasurfaces, we show that it is possible to create an anomalous non-reciprocal electromagnetic induced transparency (EIT) effect. EIT was introduced in quantum optics as a technique to enhance non-linear effects, while having strong transmission of the laser beam [26]-[27]. Its potential applications are vast, as this mechanism allows slow group velocities that can spatially compress the impinging pulse shape and enhance light-matter interactions [28]-[29]. Classical analogues of the EIT phenomenon, all satisfying reciprocity constraints, have been studied in recent years to apply these unusual wave properties to optical devices and metamaterials [30]-[32]. Here we realize a non-reciprocal EIT-like transmission response through an ultrathin metasurface characterized by transverse spatiotemporal gradients, realizing efficient coupling of light that overcomes the constraints in Eq. (1). Interestingly, at the proposed EIT peak the transmission amplitude can be made unitary, beyond the previously mentioned symmetry constraints of ultrathin surfaces, and, in addition, at the same time largely non-reciprocal, yielding, in the absence of loss, an ideal free-space isolator without forward insertion loss.

In order to demonstrate the proposed concept we consider the transmission and reflection properties of a spatiotemporally modulated metasurface lying on the $x=0$ plane and described by the following time-dependent surface impedance Lorentzian operator

$$Z_s(z,t)f = \left\{L_0 \partial_t f + C_0^{-1}[1-m\cos(\beta z - \Omega t)]\int^t f dt'\right\}, \qquad (2)$$

which models a network of distributed series inductors $L_0$ and spatiotemporally modulated capacitors $C(z,t) = C_0 + \Delta C \cos(\beta z - \Omega t)$. $\Omega, \beta$ are the temporal and spatial modulation frequencies. Expression (2) holds under the assumption of weak modulation index, i.e., $m \equiv \Delta C / C_0 \ll 1$. For the moment we neglect loss, which may be included by introducing a small series resistance, and we neglect spatial dispersion effects assuming that the surface is composed of deeply subwavelength inclusions.

For the sake of brevity, we consider only transverse-magnetic (TM) excitation, the TE solution may be found in a similar fashion yielding essentially similar results [33]. The incident magnetic field is y-polarized with longitudinal wavenumber $k_z = k\cos\theta$, $k = \omega/c$ under an $e^{-i\omega t}$ time convention, and $c$ is the speed of light. The angle $\theta$ is measured from the negative $z$ axis, as shown in Fig. 1. The reflected and transmitted fields do not need to comply with conventional Snell's law of refraction, due to the transverse gradients, and are generally written as infinite series of Floquet harmonics in both *space* and *time*: $\vec{H}^{t,r} = \hat{y} \sum_{n=-\infty}^{\infty} H_n^{t,r} e^{i(k_{z_n} z \pm k_{x_n} x - \omega_n t)} + c.c$ (see Eq. S2 in [34]), where the superscripts $t, r$ denote transmitted (reflected) fields and correspond to the upper (lower) signs; $k_{x_n} = \sqrt{k_n^2 - k_{z_n}^2}$ is the transverse wave number and, since the field satisfies the radiation condition, $\text{Im}\{k_{x_n}\} \geq 0$. The radial frequency, wavenumber, and longitudinal wavenumber of the $n$-th harmonic are $\omega_n = \omega + n\Omega$, $k_n = \omega_n/c$, $k_{z_n} = k_z + n\beta$, respectively.

Due to the electric-field continuity, the zero-th order reflected and transmitted fields propagate at the angles $\theta_r = \pi - \theta_i$ and $\theta_t = \theta_i$, respectively, as required by momentum conservation. However, this is not a fundamental constraint and it may be overcome if one uses an additional magnetic surface impedance as in [5],[8], or stacks of metasurfaces [2]. The higher-order harmonics have different transverse momentum, and exist at different frequencies than the incident wave. By further enforcing the impedance boundary condition $Z_s \hat{x} \times \left[ \vec{H} \vert_{x=0^+} - \vec{H} \vert_{x=0^-} \right] = \vec{E}_{\tan} \vert_{x=0}$, we obtain

$$A_n H_n^r - mZ_{c_{n+1}} H_{n+1}^r - mZ_{c_{n-1}} H_{n-1}^r = \delta_n H_0 \eta_0 k_{x_n} / k_n, \quad (2)$$

Where $H_n^t = -H_n^r + H_0 \delta_n$ and $\delta_n$ is the Kronecker delta, $A_n = (2Z_n + \eta_0 k_{xn}/k_n)$, $Z_n = -i\omega_n L_0 + Z_{c_n}$ and $Z_{c_n} = -1/i\omega_n C_0$. $Z_n$ and $Z_{c_n}$ are the metasurface and capacitor impedance associated with the $n$ harmonic. Eq. (2) represents an infinite set of linear equations, which, in the case of weak modulation, may be truncated to the first three harmonics $n = 0, \pm 1$.

In the absence of modulation $m=0$, the impedance is zero at the surface resonance frequency $\omega_{sr} = 1/\sqrt{L_0 C_0}$ and the surface is fully reflective, as shown in Fig 1b. When spatial modulation is introduced ($\beta \neq 0$), the surface becomes transparent in a narrow frequency band for a specified incidence direction, exhibiting an EIT-like transmission window, as shown in Fig. 1c. Yet, in the absence of temporal modulation ($\Omega = 0$), the response remains reciprocal and two full transmission peaks, corresponding to incidence angle $\theta_0$ and its complementary $\pi - \theta_0$, take place at the same frequency $\omega$, namely $T(\omega, \theta_{sr}) = T(\omega, \pi - \theta_{sr}) = 1$, as expected. Once transverse temporal modulation at frequency $\Omega$ is considered, however, reciprocity is broken, and the two resonance peaks are separated by $\bar{\omega} \sim \Omega$ (see also the discussion after Eq. (6) below), as shown in Fig. 1d, creating the opportunity for large isolation (see rays illustration there). Interestingly, as shown in the following, the bandwidth of the EIT transmission peak $\delta \omega \propto m^2$ decreases with the modulation index, resulting in larger transmission contrast for complementary directions with moderate modulation requirements. Counterintuitively, therefore, in order to enhance isolation and non-reciprocal response one has to decrease the modulation index and, for a specified modulation frequency $\Omega$, no matter how small it is, it is always possible to find a modulation index resulting in large isolation.

To prove this fact, we use Eq. (2) to solve for the reflection coefficient $R = H_0^r / H_0$ [34]

$$R^{-1} = \frac{k}{\eta k_x} \left[ A_0 - m^2 \frac{Z_{c0} Z_{c1}}{A_1} - m^2 \frac{Z_{c0} Z_{c-1}}{A_{-1}} \right]. \tag{3}$$

Interestingly, *full-transmission* and *zero coupling* to undesired diffraction orders can take place if $A_1 = 0$ or $A_{-1} = 0$. These conditions correspond to the resonant excitation of the $1, -1$ diffraction order, and they can be regarded as generalized Wood's anomalies in the case of space-time gradients. The incident wave excites a leaky-wave resonance in the structure, which completely cancels specular reflections and fully restores the incident power into the fundamental (0-th order) transmission direction [33]. As a result, a narrow transmission window is created within the angle-frequency region for which the unmodulated surface would be opaque. The resonance quality factor is proportional to the leaky mode decay rate and, in order to have full-transmission, phase matching is essential between the incident wave and the leaky mode, i.e., $k_0 \cos \theta_0 = \text{Re}\{k_z^L\}$ where $k_z^L$ is the leaky mode longitudinal wavenumber. The term "phase matching" is somewhat abused here as we discuss an enforced resonance rather than a self-sustaining one. Since $\text{Im}\{k_z^L\} \neq 0$, for strict phase matching an evanescent incident wave is required. We emphasize that the full transmission is an exact result of the infinite system of equations (2), and not an artifact of the weak modulation approximation [34].

Figures 2a and 2b show the dispersion of the real and imaginary parts of the transverse wave-vector for the 0-th and 1-st order harmonics of the TM surface and leaky modes, respectively, supported by a surface with $L_0 = \eta_0 Q / 2\omega_{SR}$ and $C_0 = 2 / \eta_0 Q \omega_{SR}$, where $Q = 10$ is the surface quality factor, $m = 0.01$, $\beta/k = 0.637$ and $\Omega/\omega_{SR} = 0.01$. For the dispersion we searched in the complex $k_x$ plane for solutions to Eq. (2) without source term. The continuous lines refer to the dispersion of physical modes which can be significantly excited by physical source while the dotted curves are weakly excited non-physical solutions of (2) [34]-[35]. Physical modes include Guided (G) Leaky-Forward (L-F) with $v_g v_p > 0$. The latter can be excited by incident plane wave that satisfies the phase matching condition. In the case of an unmodulated surface, the surface wave dispersion is symmetric and purely real, as expected, and is limited to the range $\omega > \omega_{SR}$, since TM surface waves are supported by inductive surfaces. These modes are guided along the surface, and do not couple to free-space radiation.

Spatial modulation allows coupling surface modes to radiation through higher-order harmonics, generating the EIT transparency window described above, but it still preserves the symmetry and reciprocity of the dispersion diagram. In this scenario, the dispersion diagram (see [34]) consists of an infinite set of propagation branches in both directions, shifted by $\beta$ with respect to each other. For frequencies close to $\omega_{SR}$, i.e., far from the band-gap due to periodicity, the fundamental branch of the dispersion diagram corresponding to the strongest Floquet harmonic approximately coincides with the surface wave dispersion shown in Fig. 2.

The dispersion symmetry is lifted when a temporal gradient is added, which shifts vertically the $n$-th Floquet harmonic by $n\Omega$. Then, the cut-off frequency of the leaky harmonics, which are responsible for radiation coupling, is shifted by $2\Omega$ for opposite propagation directions, as seen in Fig. 2 (blue and yellow line). As a result, with proper design it is possible to excite the leaky mode, and achieve a transparency window, from one direction but not from the complementary one. As an example, at frequency $\omega = \omega_{SR}$ (light green point in Fig 2) $k_z / k_0 = 0.3949 + 5.5 \times 10^{-4} i$, corresponding to a highly directive mode that radiates at the angle $\theta^{LW} = \cos^{-1}(0.3949) = 66.74°$, implying that an incident wave at $\theta_0 = \theta^{LW}$ would strongly couple with this resonant mode (see illustration in top Fig 2a). For this angle, Eq. (3) indeed yields full-transmission at $\omega_{SR}$. On the other hand, the symmetric point in the $\omega - \beta$ plane lies below the cutoff frequency of the leaky mode propagating in the $-z$ direction, resulting in very small transmission for a wave incident from the complementary direction $\pi - \theta_0$, and yielding isolation and strong nonreciprocal response.

The incidence angles for which full-transmission occurs can be calculated in closed-form imposing $A_1 = 0$ or $A_{-1} = 0$. In particular, assuming that $\omega \approx \omega_{sr}$ [34], we obtain the four solutions. Two are

$$\cos\theta_0 \simeq \pm\sqrt{1+\left[2(d\omega+\Omega)/\Delta\omega\right]^2} - \beta/k, \qquad (3)$$

and another two obtained by $-\Omega \mapsto \Omega$ and $-\beta \mapsto \beta$ in (3). In Eq.(3) $\Delta\omega = \omega_{sr}/Q$ is the bandwidth of the unmodulated surface, and $d\omega = \omega - \omega_{SR}$ is the frequency detuning from the resonance of the unmodulated surface. Eq. (3) is valid *if and only if* (a) either the +1 or −1 diffraction order is evanescent within the visible angular spectrum, $|k_z| < \omega/c$, i.e., $(\omega \pm \Omega)/c < |k_z \pm \beta|$, and (b) the surface impedance is inductive for the other first-order harmonic, i.e., $\omega > \omega_{SR} \mp \Omega$. The last condition is equivalent to working above the cut-off frequencies of the physical leaky modes. Eq. (3) clearly shows that, although spatial modulation is enough to achieve angularly selective transmission, it cannot break time-reversal symmetry and the constraint in (1), and the transparency window will necessarily occur at both $\theta_0$ and $\pi - \theta_0$. The only way to obtain angularly selective *non-reciprocal* transmission is by realizing a transverse spatiotemporal gradient on the surface. For the set of parameters in Fig. 2, Eq. (3) is satisfied only for $\theta_0 = 66.74°$, confirming our predictions based on the dispersion diagram in Fig. 2. For mathematical relation between full transmission condition and the dispersion diagrams in Fig. 2 see [34]. Furthermore, it is interesting to highlight that the full-transmission angle is totally independent of the modulation index $m$, which, as shown below, affects only the bandwidth of the transparency window.

Fig. 3 shows the power transmission coefficient $|T|^2 = 1 - |R|^2$ versus frequency for the incidence directions $\theta_0 = 66.74°$ and $180° - \theta_0 = 113.26°$, and the corresponding magnetic field profiles at frequency $\omega_{SR}$. The transmission coefficient was calculated analytically and numerically through Eq. (3) and Finite-Difference Time-Domain (FDTD) simulations, while the field profiles were derived through FDTD simulations. We used the same parameters as in Fig. 2, except for the modulation index which is now $m = 0.05$. Such an increase in $m$ does not affect the transmission angle or frequency, as explained above, but it reduces the EIT-like resonance Q-factor, also reducing the FDTD simulation time. For incidence at $\theta_0 = 66.74°$, the transmission becomes maximum for $\omega = \omega_{SR}$, consistent with the existence of a leaky mode at the light-green point in Fig. 2. On the other hand, for incidence at $180° - \theta_0 = 113.26°$ the transparency window is blue-shifted around $\omega = 1.02\omega_{SR}$, due to the blue-shift of the leaky mode propagating along the −z direction in Fig. 2. The field profiles in Figs. 3(b) and (c) verify that power is almost completely transmitted (reflected) for incidence from $\theta_0 = 66.74°$ ($180° - \theta_0 = 113.26°$). Additional resonances due to high diffraction orders, not seen in the weak modulation approximation, are seen in the FDTD simulation. The strong reactive fields close to the surface in Fig. 3(b) reveal the excitation of a strong resonance, which corresponds to the fundamental Floquet harmonic of the leaky mode in Fig. 2[1]. The amplitude of this harmonic can be calculated as $H_1^r = -(\eta/Z_{c_1})(k_x/k)H_0/m$, showing that, for $m \ll 1$, it can become much stronger than the incident-field amplitude $H_0$. On the other hand, in the case $180° - \theta_0 = 113.26°$ the reactive fields are very weak, since the coupling between the incident wave and the leaky mode of the surface is negligible. In such case, the impinging energy experiences specular reflection, except for a weak $n=1$ diffraction order at frequency $\omega = \omega_{SR} + \Omega$ and direction $\theta_1 = \cos^{-1}(k_z^1/k_1) = 76.1°$ with respect to $+\hat{z}$.

The anomalous EIT-like dispersion of the surface is a consequence of the interplay between two resonance phenomena: the wide resonance of the uniform metasurface and the much narrower resonance associated with the leaky mode produced by the modulation. For a specified incidence angle $\theta_0$, the EIT-resonance bandwidth and Q-factor are approximately

$$\delta\omega = m^2 Q \omega_{SR} / 4\sin\theta_0 \rightarrow Q_{FT} = 4\sin\theta_0 / m^2 Q, \qquad (4)$$

predicting an infinitely large Q-factor for infinitely small modulation index. This is related to the fact that, for weak modulation, the lifetime of the surface leaky mode increases, and becomes infinite in the limit $m \rightarrow 0$ (bound mode), when no coupling to free-space exists, opening the possibility to induce a non-reciprocal embedded scattering eigen-state on the surface [36]-[37]. Note that loss will yield a lower bound on $\delta\omega$ given by $\min\delta\omega = \Delta\omega R_o / \eta_0$ where $R_o$ is a distributed loss term [34]. The high-Q resonance allows drastic relaxation of the requirements regarding the temporal modulation frequency in order to achieve significant non-reciprocity and isolation. The frequency separation of full-transmission peaks for opposite propagation directions is $\bar{\omega} \approx \Omega + \Delta\omega\sqrt{\Omega/\omega_{SR}}$ [34]. To have isolation between the $\theta_0$ and $\pi - \theta_0$ directions we require $\bar{\omega} > \delta\omega$. Therefore, quite surprisingly, for a given modulation frequency, weaker modulation index leads to higher isolation. Eq. (4) also suggests that a lower Q-factor for the surface provides a larger resonance $Q_{FT}$. This is related to the fact that a lower surface Q implies less sensitivity to the modulation, ensuring lower energy leakage for the same modulation index. As we show in [34], the angular bandwidth also decreases as $m$ increases, following a similar square power law.

---

[1] Note that the fundamental Floquet harmonic of the leaky mode corresponds to the +1 or -1 Floquet harmonic in the case of plane-wave excitation.

A possible implementation of the surface impedance operator in Eq. (2) involves a 2D array of split ring resonators (SRR) loaded with variable capacitors, as in Fig. 4(a). This realistic structure was analyzed via full-wave finite-element simulations, with variable capacitors implemented by filling the gaps of the *n*-th row of SRRs with time-modulated dielectric material $\epsilon_r = \epsilon_r^0[1 + m\cos(\Omega t - \beta nd)]$, where $d$ is the SRR periodicity, and the inclusion geometry is provided in the caption. The modulation parameters were selected as $\beta/k = 0.793$, $m = 0.1$ and $\Omega = 0.02\omega_{SR}$. In order to relax the computational requirements of a full 3D simulation, we assumed a distance between SRRs along the *y*-direction $t < d \ll \lambda_0$, and replaced 1D arrays with a single equivalent 2D SRR, as in Fig. 4(a). The particles are made of copper taking loss effects into account. The power transmission response is given in Fig. 4(b), and it is about 85-90% due to Ohmic loss. Note that the effective modulation index is smaller than $m$ due to the discrete nature of the surface and the additional parasitic capacitances in the structure, therefore making the full-transmission bandwidth smaller than what would be expected for a uniform modulation with same amplitude. The non-reciprocal EIT-like response of the structure is evident: transmission is maximum at different frequencies for complementary incidence directions. Moreover, Fig 4b may be used to estimate the surface bandwidth $\Delta\omega \approx 0.2\omega_{SR}$ therefore $Q \approx 5$, then we approximate $\bar{\omega} \approx 0.048$ (formula given above, also in Eq. (S33) [34]) in nice agreement with the simulation. Furthermore by Eq. (3) $\theta_0 \approx 77.76°$ in fair agreement with the simulation. From Fig 4b we have $\delta\omega = 0.002\omega_{SR}$, then by substituting in Eq. (4) we obtain the effective modulation index $m_{eff} \approx 0.038$. Fig. 4(c) shows the magnetic field distribution at the maximum-transmission frequency for a $70°$ incidence angle. The left-panel refers to incidence from this direction, showing large transmission and almost zero reflection, as expected. The right panel corresponds to the complementary incident angle $110°$, for which transmission is very small.

RF implementations as proposed will typically use varactors that work up to 10GHz and have high modulation index. In optics or IR the modulation can be obtained by pn junctions, acousto optical effects, high-lasers acting on non-linear media and so on. For acoustical implementations piezo-electric components may be utilized.

To conclude, in this letter we extended the concept of graded metasurfaces by adding transverse temporal modulation of the electronic properties of a surface impedance. We showed that spatio-temporal modulation can overcome geometrical and temporal symmetry constraints of ultrathin surface, yielding non-reciprocal, angularly selective, full transmission through an ultrathin impedance surface. In our proof of concept scenario, we focused on relatively simple periodic space-time gradients, however this concept can be readily extended and applied to more sophisticated surfaces with impedance gradients that enable further control of light as in [1]-[14]. The proposed concept of space-time gratings can also be used to enhance control over near-fields, and to create non-reciprocal radiation[38]-[39], opening new venues for efficient source-field manipulation.

**Figures**

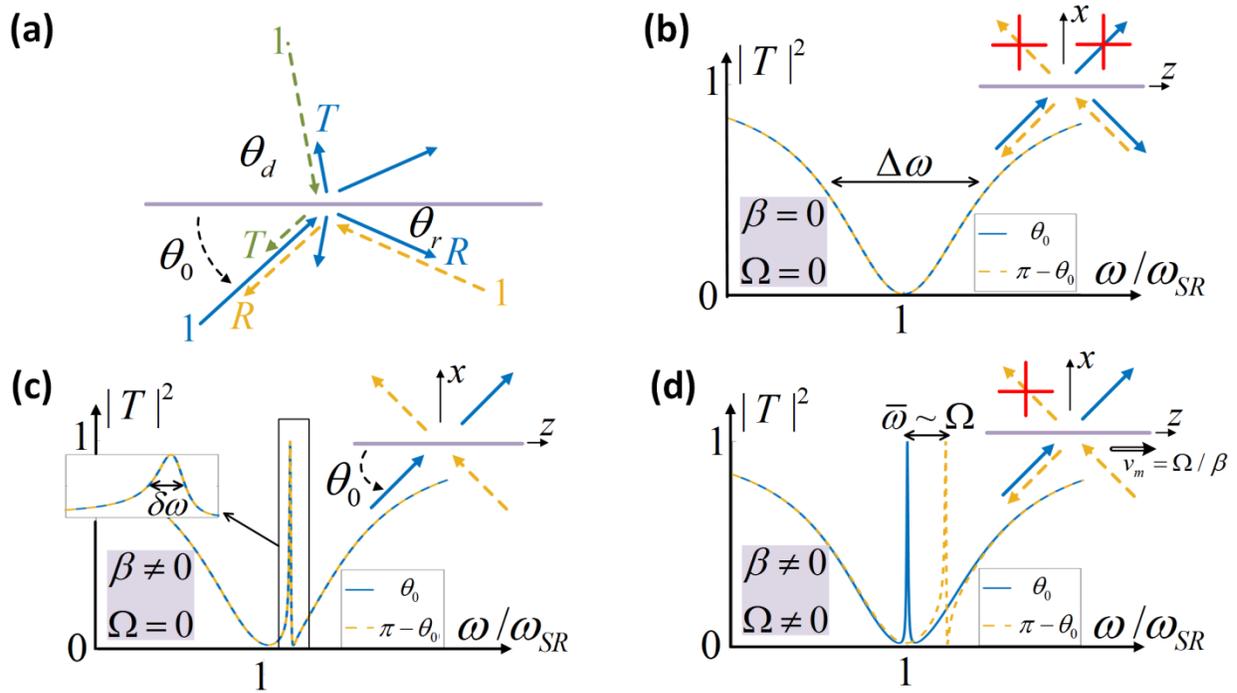

**Figure 1. Illustration of the main concept**. The purple horizontal lines represent the surface impedance. **(a)** The constraints of a conventional reciprocal surface with spatial gradients. **(b)** Typical transmission of the surface impedance (2) with no modulation. **(c)** Spatial modulation only can produce a reciprocal EIT-like transmission as will be discussed in the following **(d)** Spatiotemporal modulation can provide isolation. The asymmetry of the $\pi - \theta_0$ line shape is due to a Fano-like interaction between the two resonances we have in this system.

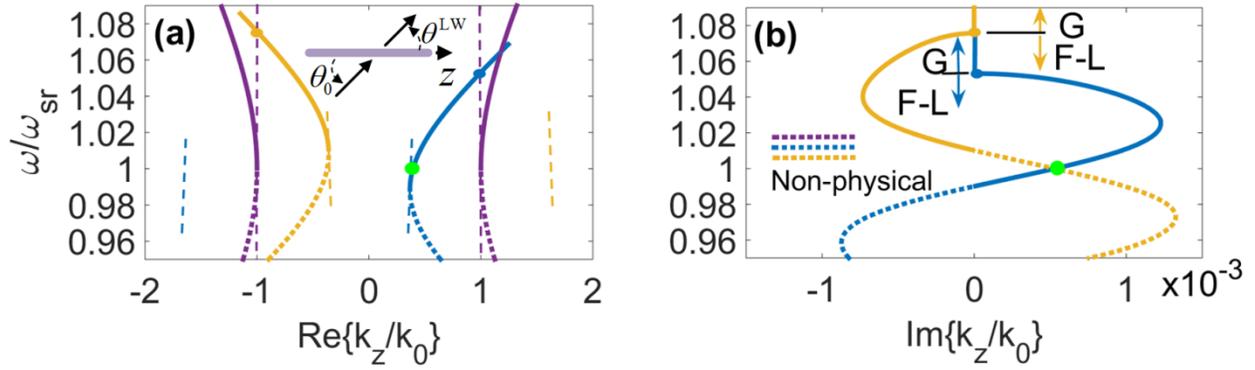

**Figure 2. Dispersion of the surface TM modes. (a)** Continuous lines describe the dispersion of physical waves, dotted lines describe the dispersion of non-physical waves that cannot be excited. In purple the dispersion of TM modes on unmodulated surface impedance. Blue and yellow curves are obtained for a spatiotemporally modulated surface with $\beta/k = 0.637$, $\Omega/\omega_{sr} = 0.01$, $m = 0.01$. In purple, yellow, and blue dashed lines we mark the boundaries of the light cone of the n=0, n=1 and n=-1 harmonics, respectively. **(b)** Imaginary part of the modes wave number. The light-green circle indicates the operation point for the results shown in Fig. 3. The inset in (a) illustrates the enforced excitation of a leaky mode by an incident plane wave under the phase matching condition.

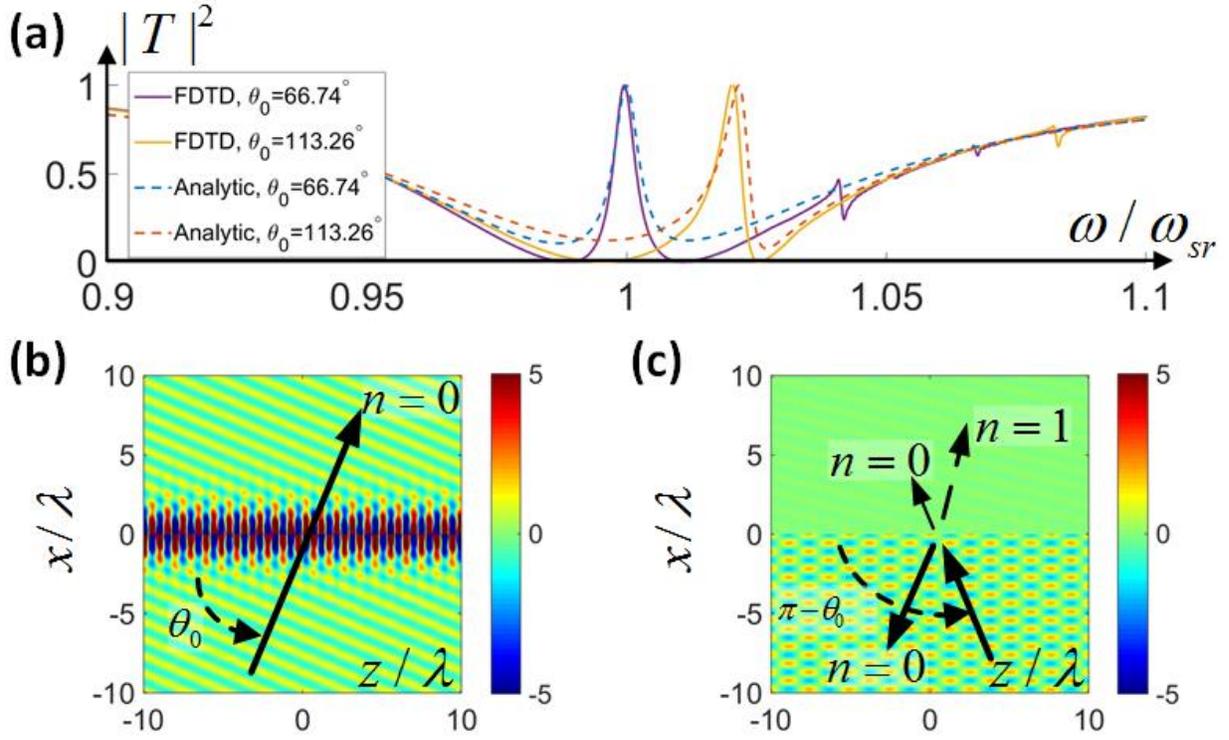

**Figure 3. Non-reciprocal surface response. (a)** Comparison between the transmission vs frequency for two complimentary incident waves at $\theta_0 = 66.74°$ and $180° - \theta_0 = 113.26°$. The response was calculated numerically by FDTD code and analytically. Simulations carried out with $\Omega/\omega_{sr} = 0.01$, $m = 0.05$, $\beta/k_0 = 0.637$. This simulation demonstrates the accuracy of the weak modulation approximation we use for the analytic results given in this text **(b)** At $\omega = \omega_{SR}$ and with $\theta_0 = 66.74°$ full transmission takes place. The reactive power near the surface is high due to the enforced excitation of a weakly radiating leaky mode. No other propagating diffraction order is excited. **(c)** At the same frequency, $\omega = \omega_{SR}$, but with complementary angle of incidence $180° - \theta_0$, no leaky mode is excited, the surface is practically opaque. The n=1 harmonic is weakly excited at a different frequency than the incident wave.

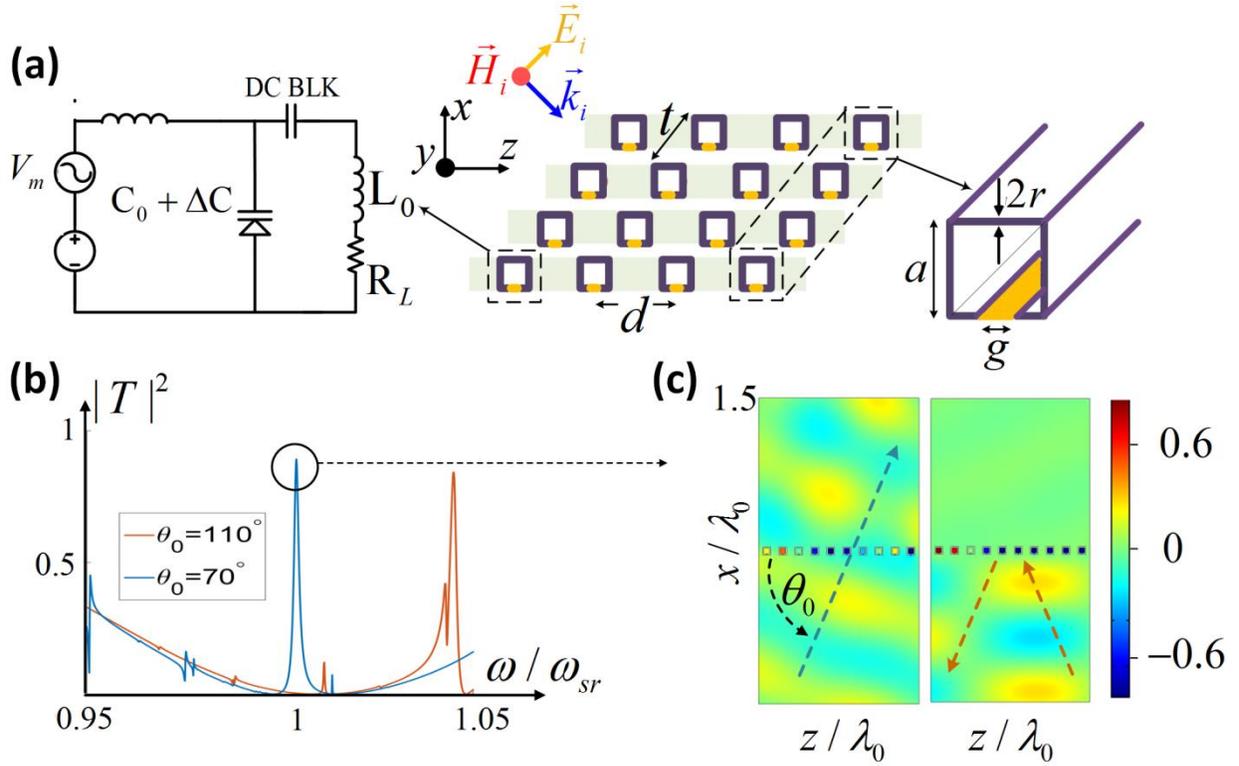

**Figure 4. Implementation using an array of split ring resonators. (a)** Geometry of the array. Left inset: lumped circuit model for the loaded loop and biasing network. $L_0, R_L$ are the equivalent inductance, and radiation/Ohmic resistance of a single loop. Right inset: zoom on a single loop as implemented in the finite-element simulation. The gap is filled with a time-varying dielectric (yellow). The side length, gap and metal thickness of the SRRs were selected as $a = \lambda_0/15$, $g = 0.0046\lambda_0$ and $r = 0.01\lambda_0$, respectively, with $\lambda_0$ the resonance wavelength. The size of the structure along the z-direction is $D = 2\pi/\beta$, and the periodicity of the lattice is $d = D/N$, with $N = 10$. **(b)** Transmission response at two complementary angles. The transmission peaks are around 90% due to the Ohmic loss in the copper particles. The additional transmission spikes are due to higher order harmonics neglected in the weak modulation approximation (see discussion regarding Fig. 3 as well) **(c)** Fields picture at $\omega/\omega_{SR} \approx 1$. Almost full transmission is obtained at $\theta_0 = 70°$, as opposed to the high isolation obtained at $\theta_0 = 110°$.

**References**

[1] N. Yu, P. Genevet, M. A. Kats, F. Aieta, J.-P. Tetienne, F. Capasso, and Z. Gaburro, "Light propagation with phase discontinuities: Generalazed laws of reflection and refraction" Science **334** 6054 333-337 (2011)

[2] F. Monticone, N. M. Estakhri, and A. Alu, "Full control of nanoscale optical transmission with a composite metascreen", *Phys. Rev. Lett* **110** 203903 (2013)

[3] Y. Li, X. Jiang, R. Q. Li, B. Liang, X. Y. Zou, L. L. Yin, and J. C. Cheng, "Experimental realization of full control of refleted wave with subwavelength acoustic metasurfaces", arxiv 1407.1138 (2014)

[4] K. Sarabandi and N. Behdad, "A frequency selective surface with miniaturized elements", *IEEE Trans. Ant. Prop*. **55** 5 (2007)

[5] C. Pfeiffer and A. Grbic, "Millimeter-wave transmitarrays for wavefront and polarization control", IEEE Trans. Ant. Prop. **61** 12 (2013)

[6] M. Selvanayagam and G. V. Eleftheriades, "Circuit modelling of Huygens surfaces", *IEEE Antennas and wireless Prop. Lett.* **12** (2013)

[7] M. Selvanayagam and G. V. Eleftheriades, "Discontinuous electromagnetic fields using orthogonal electric and magnetic currents for wavefront manipulation", *Opt. Exp*. **21** 12 (2013)

[8] C. Pfeiffer and A. Grbic, "Metamaterial Huygens' surfaces: tayloring wave fronts with reflectionless sheets", *Phys Rev Lett* **110** 197401 (2013)

[9] D. Sievenpiper, J. Schaffiner, R. Loo, G. Tangonan, S. Ontiveros, and R. Harold, "a tunable impedance surface performing as a reconfigurable beam steering reflector", *IEEE Trans. Ant. Prop* **50** 3 (2002)

[10] E. Hasman, V. Kleoner, G. Biener, and A. Niv, "Polarization dependent focusing lens by use of quantized Pancharatnam – Berry phase diffractive optics", *App. Phys. Lett.* **82** 3 (2003)

[11] Dianmin Lin, Pengyu Fan, Erez Hasman, and Mark L. Brongersma, "Dielectric gradient metasurface optical elements", Science **345** 298 (2014)

[12] O. Avayu, O. Eisenbach, R. Ditcovski, and T. Ellenbogen, "Optical Metasurfaces for Polarization Controlled Beam Shaping", to be published in *Optics Letters* (2014)

[13] C. L. Holloway, M. A. Mohamed, E. F. Kuester, and A. Dienstfrey, "Reflection and transmission properties of metafilm: With an application to controllable surface composed of resonant particles", *IEEE Trans. Electromagnetic Comp.* **47** 4 (2005)

[14] E. F. Kuester, M. A. Mohamed, M. Piket-May and C. L. Holloway, "Averaged transition conditions for electromagnetic fields at a metafilm", *IEEE Trans. Ant. Prop*. **51** 10 (2003)

[15] N. Meinzer, W. L. Barnes, and I. R. Hooper, "Plasmonic meta-atoms and metasurfaces", *Nat. Photonics*, 8, 889-898 (2014)

[16] V. S. Asadchy, Y. Ra'di, J. Vehmas, and S. A. Tretyakov, "Functional metamirrors Using Bianisotropic Elements", *Phys. Rev. Lett.* **114** 095503 (2015).

[17] Y. Mazor and B. Z. Steinberg, "Metaweaves: Sector-way nonreciprocal metasurfaces" *Phys. Rev. Lett.* **112** 153901 (2014).

[18] A. M. Mahmoud, A. R. Davoyan, and N. Engheta, "All-Passive nonreciprocal metasurface", *arxiv* 1407-1812 (2014)

[19] Z. Yu and S. Fan, "Complete optical isolation created by indirect interband phtonic transitions", *Nat. Photonics*, **3** 10.1038 (2009)

[20] K. Fang and S. Fan, "Control the flow of light using inhomogenous effective gauge field that emerges from dynamic modulation", *Phys. Rev. Lett.*, **111**, 203901 (2013)

[21] Q. Lin and S. Fan, "Light guiding by effective gauge field for photons", *Phys. Rev. X*, **4** 031031 (2014)

[22] D. L. Sounas, C. Caloz, A. Alu, "Giant non-reciprocity at the subwavelength scale using angular momentum-biased metamaterials", *Nat. Phot.* 10.1038 (2013)

[23] R. Fleury, D. L. Sounas, C. F. Sieck, M. R. Haberman, and A. Alu, "Sound isolation and giant linear nonreciprocity in a compact acoustic circulator", *Science* **343** 516 (2014)

[24] D. L. Sounas and A. Alu, "Angular momentum biased nanorings to realize magnetic free integrated optical isolation", *ACS Photonics*, **1** 198-204 (2014)

[25] A. Bahabad, "Diffraction from moving grating", *Opt. Quant. Electron* 46 1065-1077 (2014)

[26] S. E. Harris, J. E Field, and A. Imamoglu, "Nonlinear Optical Processes using Electromagnetically Induced Transparency", *Phys. Rev. Lett,* **64** 10 pp.1107-1110 (1990)

[27] K. –J. Boller, A. Imamoglu, and S. E. Harris, "Observation of Electromagnetically Induced Transparency", *Phys. Rev. Lett.* **66** 20 pp. 2593-2596 (1991)

[28]     C. Liu, Z. Dutton, C. H. Behroozi, and L. V. Hau, "Observation of coherent optical information storage in an atomic medium using halted light pulses", *Nature* **409** pp. 490-493 (2001)

[29]     D. Budker, D. F. Kimball, S. M. Rochester, and V. V. Yashchuk, "Nonlinear Magneto-optics and Reduced Group Velocity of Light in Atomic Vapor with Slow Ground State Relaxation", *Phys. Rev. Lett.* **83** 9 pp. 1767-1770 (1999)

[30]     M. F. Yanik, W. Suh, Z. Wang, and S. Fan, "Stopping Light in a Waveguide with an All-Optical Analog of Electromagnetically Induced Transparency," *Phys. Rev. Lett*. **93** 233903 (2004)

[31]     Q. Xu, S. Sandhu, M. L. Povinelli, J. Shakya, S. Fan, and M. Lipson, "Experimental realization of an On-Chip All Optical Analogue to Electromagnetically Induced Transparency", *Phys. Rev. Lett.,* **96** 123901 (2006)

[32]     N. Papasimakis, V. A. Fedotov, and N. I. Zheludev, "Metamaterial Analog of Electromagnetic Induced Transparency", *Phys. Rev. Lett.* **101** 253903 (2008)

[33]     A. Hessel and A. A. Oliner, "A new theory of wood's anomalies on optical grating", *Applied Optics* **4** 10 1275-1297 (1965)

[34]     The Supplementary Online Material reports the detailed derivations of some of the equations in the paper.

[35]     F. Capolino, Theory and phenomena of metamaterials, CRC Press, Taylor & Francis group, Boca-Raton FL (2009)

[36]     C.W. Hsu, B. Zhen, J. Lee, S.-L. Chua, S. G. Johnson, J. D. Joannopoulos, and M. Soljačić, "Observation of trapped light within the radiation continuum", *Nature* **499**, 188 (2013).

[37]     F. Monticone, and A. Alù, "Embedded Photonic Eigenvalues in 3D Nanostructures," *Phys. Rev. Lett.* **112**, 213903 (2014).

[38]     Y. Hadad and B. Z. Steinberg, "One-way optical waveguides for matched non-reciprocal nanoantennas with dynamic beam scaning functionality", *Opt. Exp.* **21** S1 A77-A83 (2012)

[39]     Y. Hadad, Y. Mazor, and Ben Z. Steinberg, "Green's function theory for one-way particle chains", *Phys. Rev. B* **87** 035130 (2013).